\documentclass{article}
\usepackage{spconf,amsmath,graphicx}
\usepackage{epsfig,amssymb,amsmath,multirow,comment}
\title{Supervised Speaker Embedding De-Mixing in Two-Speaker Environment}
\name{Yanpei Shi, Thomas Hain}
\address{ Speech and Hearing Research Group\\
		Department of Computer Science, University of Sheffield\\
	\texttt{\{YShi30, t.hain\}@sheffield.ac.uk}}
\begin{document}
%\ninept
%
\maketitle
\begin{abstract}
Separating different speaker properties from a multi-speaker environment is challenging. Instead of separating a two-speaker signal in signal space like speech source separation, a speaker embedding de-mixing approach is proposed. The proposed approach separates different speaker properties from a two-speaker signal in embedding space. The proposed approach contains two steps. In step one, the clean speaker embeddings are learned and collected by a residual TDNN based network. In step two, the two-speaker signal and the embedding of one of the speakers are both input to a speaker embedding de-mixing network. The de-mixing network is trained to generate the embedding of the other speaker by reconstruction loss. Speaker identification accuracy and the cosine similarity score between the clean embeddings and the de-mixed embeddings are used to evaluate the quality of the obtained embeddings. Experiments are done in two kind of data: artificial augmented two-speaker data (TIMIT) and real world recording of two-speaker data (MC-WSJ). Six different speaker embedding de-mixing architectures are investigated. Comparing with the performance on the clean speaker embeddings, the obtained results show that one of the proposed architectures obtained close performance, reaching 96.9\% identification accuracy and 0.89 cosine similarity.

\end{abstract}

\vspace*{-2mm}
\section{Introduction}
\vspace*{-2mm}
In recent years, speech source separation becomes an active research area. Speech source separation separates mixture speech signal in signal space. Traditionally, speech source separation is viewed as a signal processing problem, different approaches are proposed such as CASA \cite{bregman1990auditory}. Matrix factorization methods are also widely used in speech source separation, such as Non-Negative Matrix Factorization (NMF) \cite{wilson2008speech,raj2010non} and Independent component analysis (ICA) \cite{saruwatari2003blind,casey2000separation,chien2006new,lee2005blind,makino2007frequency}. 
With the rapid growth of deep learning, some deep learning approaches are used to separate speech signals, such as supervised separation \cite{huang2014deep,du2014speech,tu2014speech,du2016regression}, deep clustering and deep attractor network \cite{hershey2016deep,chen2017deep,luo2018speaker}.

However, separating speech signal from two-speaker signal is still a challenging task. Speech signals are high dimensional, and different speaker properties in two-speaker signals are highly co-related to each other, which would influence the quality of the output \cite{luo2018speaker,lee2005blind}.

Instead of separating speech signal in signal space, de-mixing different speaker properties from two-speaker signal in embedding space might be more efficient. Speaker embedding is low dimensional, and it can project variable length acoustic signal into fixed length embedding space \cite{chung2018speech2vec}. This property of speaker embedding makes it convenient to be further used comparing with that in signal space. The obtained speaker embeddings might be beneficial for downstream tasks such as speaker identification \cite{snyder2019speaker,kamper2016deep,snyder2016deep,okabe2018attentive} and speech recognition \cite{watada2016speech,palaskar2019learned}.

In this work, a speaker embedding de-mixing approach for separating speaker embeddings in two-speaker signal is proposed.
The proposed approach contains two steps: in step one, a residual TDNN network is used to learn high quality speaker embeddings from clean speech data. After training, the embedding of each speaker are extracted and collected. In step two, a speaker embedding de-mixing network is trained. Suppose the input data contains a target speaker and an interfering speaker. The proposed approach takes the two-speaker signal as input, as well as the embedding of the interfering speaker. The output would be the embedding of the target speaker; or inversely, the proposed approach takes the two-speaker mixture signal and the embedding of the target speaker as input, the obtained embedding would be the embedding of the interfering speakers. When the embedding of one of the speaker is available, the system will generate the embedding of the other speaker that appears in the input signal.

To the best of our knowledge, the proposed approach is the first that trying to directly de-mix speaker embedding from two-speaker signal. This is also the main contribution of this work. The benefits of the proposed approach is manifold: Suppose in a home device, the embedding of some speakers might be available. The proposed approach might be beneficial for obtaining the embedding of the other speaker in two-speaker signal. The de-mixed speaker embedding might be further used for some downstream tasks, such as speaker verification \cite{chung2018voxceleb2,xie2019utterance} and speech recognition \cite{denisov2019end}.

The rest of this report is organized as follow: Section \ref{Model Architecture} introduces the model architectures in both step one and step two. Section \ref{Experiments} introduces the experiments design, including data and use, and experiment setup. In Section \ref{Results and Discussion}, results are shown, followed by discussion and analysis. Section \ref{Conclusion and Future Work} introduces the conclusion and the future work plan.

\vspace*{-3mm}
\section{Model Architecture}\label{Model Architecture}
\vspace*{-3mm}
In this section, the model structure in this work is introduced, which consists of two steps. Step one: learning clean speaker representation; Step two: using the learned speaker embedding to train a speaker embedding de-mixing network. The goal for step one is to learn high quality embeddings for each speaker in the dataset. In step two, the two-speaker signal is firstly projected into embedding space, resulting in mixture embedding $\boldsymbol e_{mix}$. The mixture embedding and the embedding of one of the speakers $\boldsymbol e_{2}$ are put into a de-mixing function. The output is the estimation of the embedding from the other speaker $\boldsymbol e_{1}^{'}$. 
\vspace*{-3mm}
\subsection{Step One: Learning High Quality Speaker Representations}
\vspace*{-3mm}
\begin{comment}
\begin{figure}[h]
	\centering
	\includegraphics[height=4.3cm,width=8cm]{step1.png}
	\caption{The diagram of step one. $\boldsymbol A$ is noted as the residual TDNN based speaker embedding extractor. $\boldsymbol B$ is denoted as the speaker embedding classifier.}
	\label{step1}
\end{figure}
\end{comment}

In step one, the clean speech signal is input to a speaker embedding extractor $\boldsymbol A$. After training $\boldsymbol A$, the embedding for each speaker is extracted from the bottleneck layer of $\boldsymbol A$. A classifier $\boldsymbol B$ is used to evaluate the quality of the learned speaker embeddings. 

\begin{figure}[h]
	\centering
	\includegraphics[height=8cm,width=8cm]{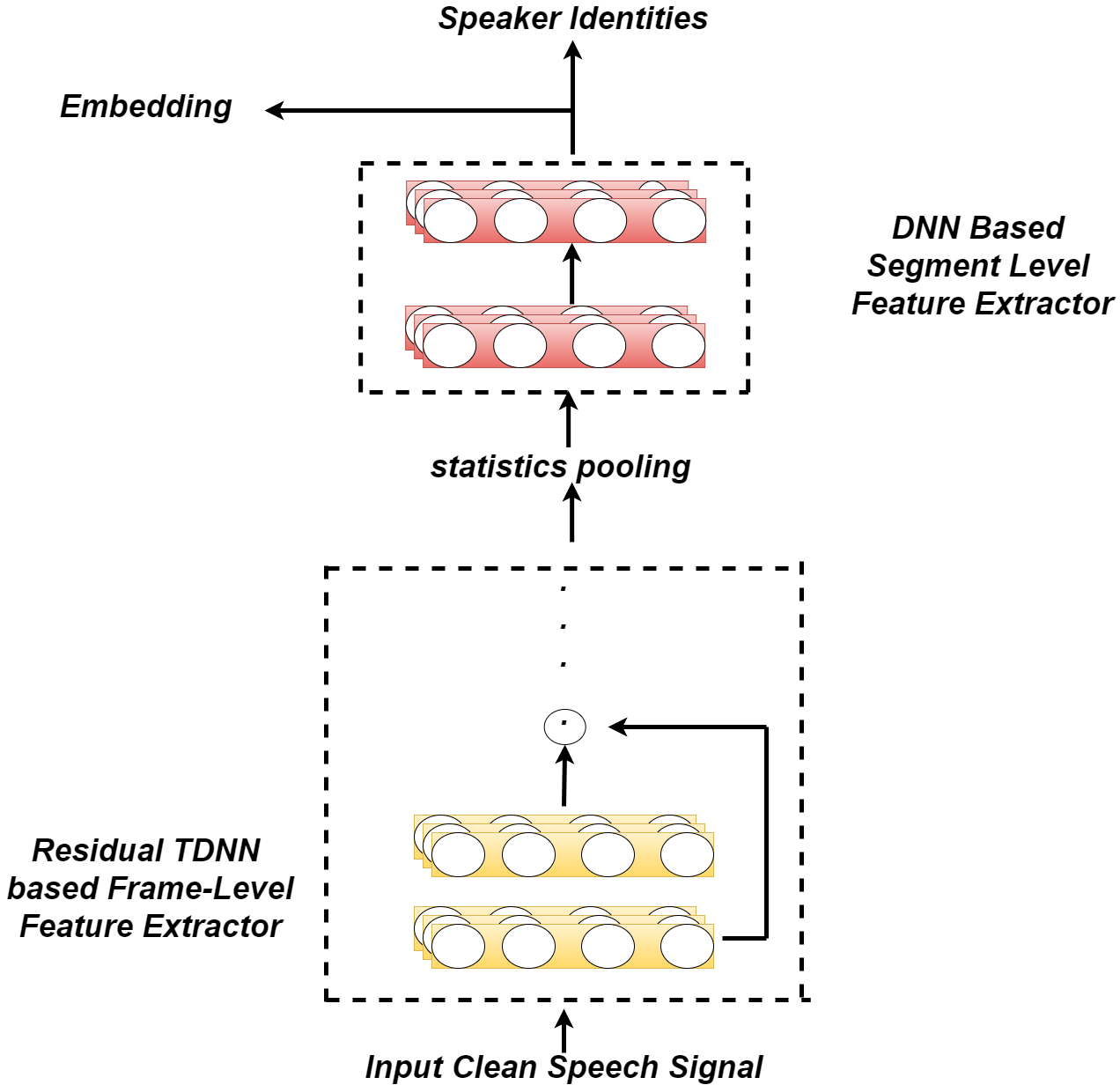}
	\vspace*{-3mm}
	\caption{The architecture of speaker embedding extractor $\boldsymbol A$.}
	\label{xvector}
\end{figure}

\begin{comment}
\begin{figure}[h]
	\centering
	\includegraphics[height=5cm,width=8cm]{step2.png}
	\caption{The diagram of Step Two. $\boldsymbol C$ is the speaker embedding de-mixing network. $\boldsymbol B$ is the fixed speaker embedding classifier that trained on step one.}
	\label{step2}
\end{figure}
\end{comment}

\begin{figure}[h]
	\centering
	\includegraphics[height=10cm,width=8cm]{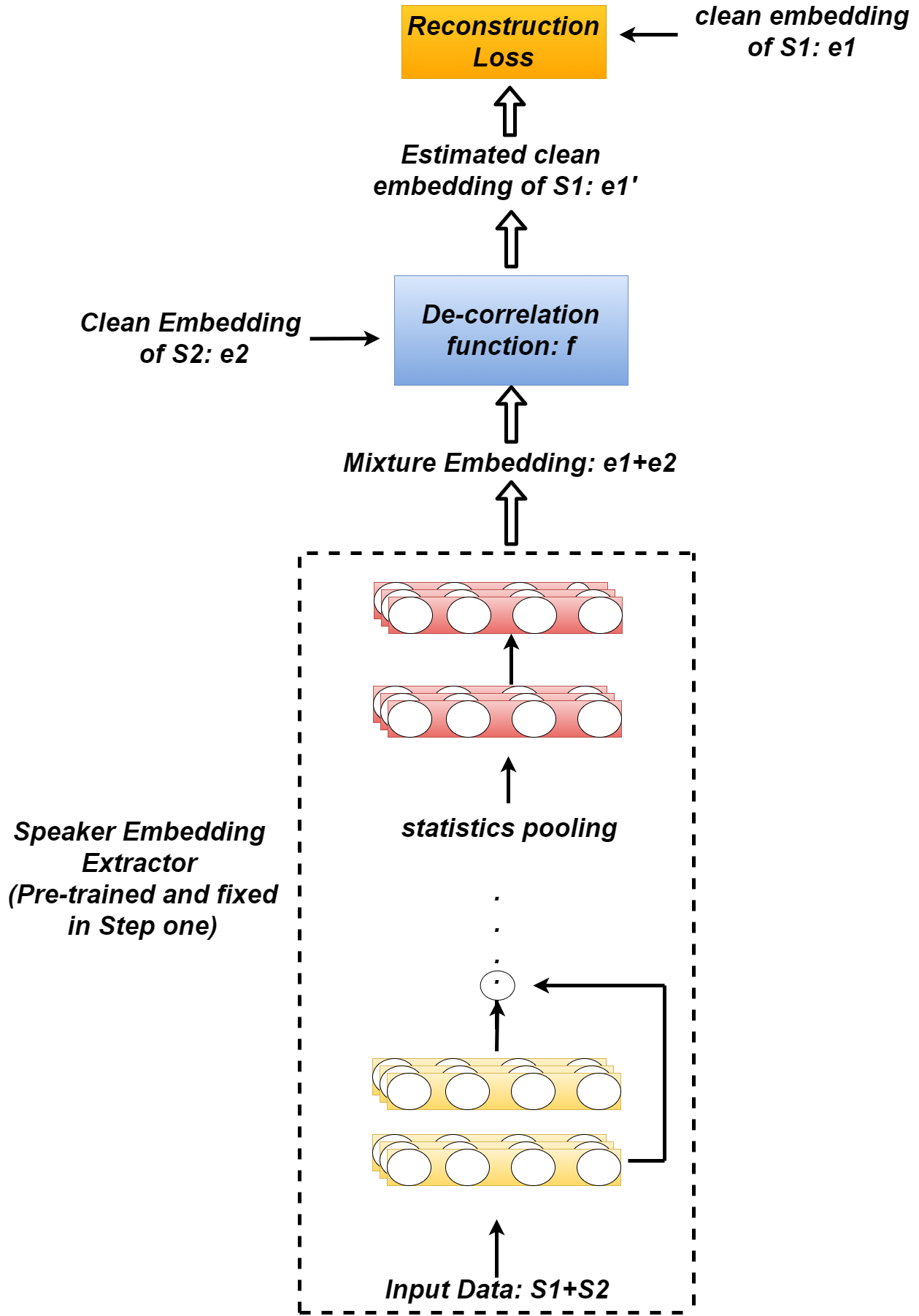}
	\vspace*{-3mm}
	\caption{Model Architecture of speaker embedding de-mixing network $\boldsymbol C$. $\boldsymbol C$ consists of pre-trained speaker embedding extractor and the de-mixing function.}
	\label{multi-net}
\end{figure}

\begin{figure*}[h]
	\centering
	\includegraphics[height=11cm,width=16cm]{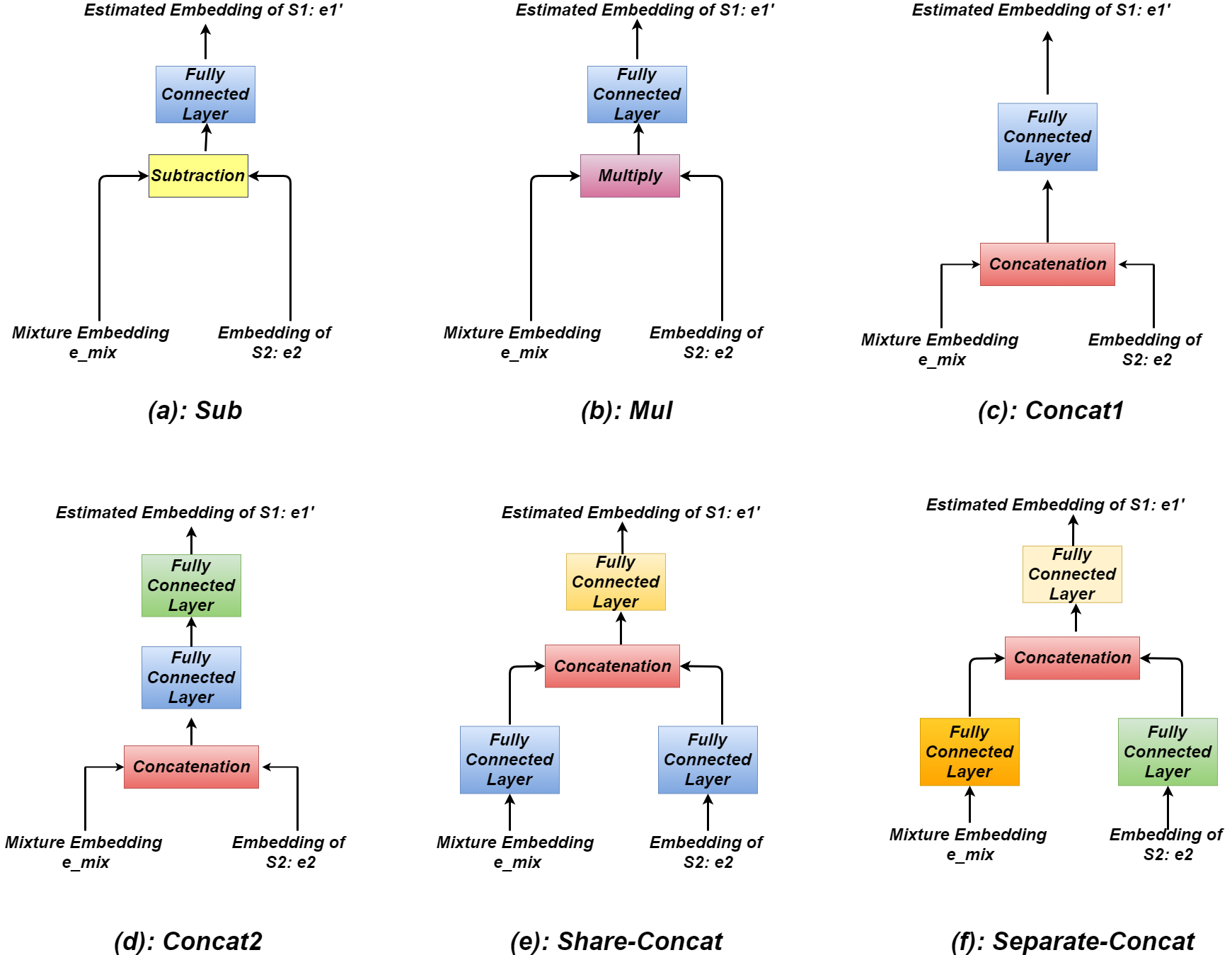}
	\vspace*{-3mm}
	\caption{Different architecture of de-mixing function $\boldsymbol f$: (a) Subtraction; (b) Multiplication; (c) Concatenation with one fully-connected layer (d) Concatenate with two fully-connected layers; (e) Shared Fully-Connected Layer with Concatenation and (f) Separated Fully-Connected Layer with Concatenation.}
	\label{choice}
\end{figure*}

\begin{table}[h]
	\renewcommand\arraystretch{1.0}
	\setlength{\tabcolsep}{2mm}
	\centering  
	\footnotesize
	\begin{tabular}{c|c|c}
		\hline
		Layer& Context & Output  \\
		\hline
		TDNN Layer1&$[t-1, t, t+1]$& 512\\
		\hline
		TDNN Layer2&[t]& 512\\
		
		\hline
		\multirow{2}{*}{TDNN-Res1}&$[t-2,t-1,t,t+1 t+2]$&\multirow{2}{*}{512}\\
		&[t]\\
		\hline
		
		\multirow{2}{*}{TDNN-Res2}&$[t-2,t-1,t,t+1, t+2]$&\multirow{2}{*}{512}\\
		&[t]\\
		\hline
		\multirow{2}{*}{TDNN-Res3}&$[t-2,t-1,t,t+1, t+2]$&\multirow{2}{*}{512}\\
		&[t]\\
		\hline
		TDNN Layer3&[t]& 1500\\
		\hline
		Statistics Pooling &T&3000\\
		\hline
		\multirow{2}{*}{Segment-Level}&
		T & 512\\
		&T& 512\\
		\hline
	\end{tabular}
\vspace*{-2mm}
	\caption{Architecture of the speaker embedding network $\boldsymbol A$}\label{model_summary}
	\label{model_sum}
\end{table}

Figure \ref{xvector} and Table \ref{model_sum} show the architecture of $\boldsymbol A$. In order to learn high quality and robust speaker embeddings, $\boldsymbol A$ is designed based on TDNN architecture, as TDNN architecture shows high robustness and it can better capture time relevant information \cite{snyder2018x}.
There are three parts within the architecture of $\boldsymbol A$: frame-level feature extractor, statistics pooling and segment-level feature extractor. 

In frame-level feature extractor, the network consists of TDNN layers and residual TDNN blocks. The input data is firstly passed through into two TDNN layers. Then, three residual TDNN blocks are used. The last TDNN layer transforms the feature dimension into 1500. The use of residual TDNN blocks instead of using normal TDNN layers like X-vectors might increase the robustness of the learned embeddings \cite{zeinali2019but}. 

Statistics pooling operation is then used, the output is feed into the segment-level feature extractor. There are two fully-connected layers in segment-level feature extractor. The speaker embedding is extracted from the last fully-connected layer.

For the architecture of classifier $\boldsymbol B$, a simply architecture is chosen: a fully connected network with one hidden layer with 512 nodes.

\vspace*{-3mm}
\subsection{Step Two: The De-mixing of Speaker Representations in Embedding Space}
\vspace*{-2mm}

After collecting the high quality embeddings for each speaker in step one, step two learns the de-mixing function of the mixture embeddings. 

Suppose the input data contains two speakers: $\boldsymbol s_{1}$ and $\boldsymbol s_{2}$. In step one, both of the high quality embeddings of $\boldsymbol s_{1}$ and $\boldsymbol s_{2}$ are learned and obtained, which are denoted as $\boldsymbol e_{1}$ and $\boldsymbol e_{2}$. Given the input mixture data, the speaker de-mixing network $\boldsymbol C$ firstly transforms it in embedding space, results in mixture embedding $\boldsymbol e_{mix}$. Then, a de-mixing function $\boldsymbol f$ is learned to remove the information of the speakers and remains the other.

More specifically, Figure \ref{multi-net} illustrates the architecture of de-mixing network $\boldsymbol C$. The input mixture data contains $\boldsymbol s_{1}$ and $\boldsymbol s_{2}$. $\boldsymbol C$ contains two-parts: the first part contains the pre-trained speaker embedding extractor in step one, the goal is to project the input data in embedding space. The output of the pre-trained speaker embedding extractor is $\boldsymbol e_{mix}$, which consists of the mixture embedding of two speakers: $\boldsymbol e_{1} + \boldsymbol e_{2}$. Then, $\boldsymbol e_{mix}$ and the clean embedding $\boldsymbol e2$ (trained and collected from step one) are input to a de-mixing function $\boldsymbol f$ (shows in Equation \ref{f}). The output is estimated embedding of the other speaker $\boldsymbol e1^{'}$.

\begin{equation}\label{f}
\boldsymbol e_{1}^{'} = \boldsymbol f(\boldsymbol e_{mix},\boldsymbol e_{2})
\end{equation}

A reconstruction loss $\mathcal{L}$ (shows in Equation \ref{recon}) is applied between $\boldsymbol e_{1}^{'}$ and $\boldsymbol e_{1}$. In this work, mean absolute error \cite{willmott2005advantages} is applied. 

\begin{equation}\label{recon}
\mathcal{L}= ||\boldsymbol e_{1}-\boldsymbol e_{1}^{'}||
\end{equation}

\vspace*{-2mm}
\subsection{The architecture of the de-mixing function $\boldsymbol f$}
\vspace*{-2mm}
The de-mixing function $\boldsymbol f$ might have different choices. In this work, six possible methods are investigated. Figure \ref{choice} illustrates the six different methods of $\boldsymbol f$: (a) Subtraction; (b) Multiplication; (c) Concatenation with one fully-connected layer (d) Concatenate with two fully-connected layers; (e) Shared Fully-Connected Layer with Concatenation and (f) Separated Fully-Connected Layer with Concatenation. 
\vspace*{-4mm}
\subsubsection{Subtraction}
\vspace*{-2mm}
The first one is a subtraction operation of $\boldsymbol e_{mix}$ and $\boldsymbol e_{2}$ (shows is Equation \ref{sub} and Figure \ref{choice} (a)). After subtraction, the subtracted embedding vector is passed through a fully-connected layer without activation function (could be viewed as a linear transformation). This method is further referred to ``Sub". The embedding dimension is denoted as $d$. $\boldsymbol W \in \mathcal {R}^{d \times d}$ and $\boldsymbol b \in \mathcal {R}^{1 \times d}$ are the parameters of the fully-connected layer. 

\begin{equation}\label{sub}
\boldsymbol f(\boldsymbol e_{mix},\boldsymbol e_{2}) = (\boldsymbol e_{mix} - \boldsymbol e_{2})\boldsymbol W+ \boldsymbol b
\end{equation}
\vspace*{-11mm}
\subsubsection{Multiplication}
\vspace*{-2mm}
Multiplication approach (further referred to ``Mul") is similar with ``Sub" method. The only difference is $\boldsymbol e_{mix}$ is multiplied with $\boldsymbol e_{2}$ instead of subtracted. Figure \ref{choice} (b) and Equation \ref{mul} shows the architecture of ``Mul" method. $\odot$ denotes element-wise multiplication.

\begin{equation}\label{mul}
\boldsymbol f(\boldsymbol e_{mix},\boldsymbol e_{2}) = (\boldsymbol e_{mix} \odot \boldsymbol e_{2})\boldsymbol W+\boldsymbol b
\end{equation}
\vspace*{-5mm}
\subsubsection{Concatenate with one fully-connected layer}
\vspace*{-2mm}
In the third method, $\boldsymbol e_{mix}$ and $\boldsymbol e_{2}$ are firstly concatenated together, and then feeded into a fully connected layer (shows in Equation \ref{concat1} and Figure \ref{choice} (c)). $[\boldsymbol e_{mix},\boldsymbol e_{2}]^{T} \in \mathcal {R}^{1 \times 2d}$ denotes the concatenated vector of $\boldsymbol e_{mix}$ and $\boldsymbol e_{2}$. This method is further referred to ``Concat1". $\boldsymbol W \in \mathcal {R}^{2d \times d}$ and $\boldsymbol b \in \mathcal {R}^{1 \times d}$ are parameters for the fully connected layer, $\times$ denotes matrix multiplication.

\begin{equation}\label{concat1}
\begin{aligned}
\boldsymbol f(\boldsymbol {e_{mix}},\boldsymbol {e_{2}}) &= [\boldsymbol e_{mix},\boldsymbol e_{2}]^{T} \times \boldsymbol W+\boldsymbol b\\
\end{aligned}
\end{equation}

\vspace*{-5mm}
\subsubsection{Concatenate with two fully-connected layers}
\vspace*{-2mm}

The next method is concatenate with two fully-connected layers. Similar with the previous method,  $\boldsymbol e_{mix}$ and $\boldsymbol e_{2}$ are firstly concatenated together, and then feed into two fully connected layers instead of one (shows in Equation \ref{concat2} and Figure \ref{choice} (d)). The first fully-connected layer 
uses $Relu$ activation function while there are no activation function after the second layer. 

This method is further referred to ``Concat2". $\boldsymbol W \in \mathcal {R}^{2d \times d}$ and $\boldsymbol b \in \mathcal {R}^{1 \times d}$ are parameters for the fully connected layer. 

\begin{equation}\label{concat2}
\begin{aligned}
\boldsymbol f(\boldsymbol {e_{mix}},\boldsymbol {e_{2}}) &= \mathrm {Relu}(([\boldsymbol e_{mix},\boldsymbol e_{2}]^{T} \times \boldsymbol W_{0}+\boldsymbol b_{0})\boldsymbol W_{1})\\
\end{aligned}
\end{equation}

\vspace*{-5mm}
\subsubsection{Shared Fully-Connected Layer with Concatenation}
\vspace*{-2mm}
The last two methods are different from the above methods. In the fifth method,  $\boldsymbol e_{mix}$ and $\boldsymbol e_{2}$ are firstly input to two fully connected layers respectively, the two fully connected layer share parameters. The output $\boldsymbol k_{mix}$ and $\boldsymbol k_{2}$ are then concatenated and feed into another fully connected layer (shows in Equation \ref{share_concat} and Figure \ref{choice} (e)). This method is further referred to ``Share-Concat". $\boldsymbol W_{0} \in \mathcal {R}^{d \times d}$, $\boldsymbol b_{0} \in \mathcal {R}^{1 \times d}$, $\boldsymbol W_{1} \in \mathcal {R}^{2d \times d}$ and $\boldsymbol b_{1} \in \mathcal {R}^{1 \times d}$ are parameters for the fully connected layers.

\begin{equation}\label{share_concat}
\begin{aligned}
\boldsymbol f(\boldsymbol e_{mix},\boldsymbol e_{2}) &= \mathrm {Relu}([\boldsymbol k_{mix},\boldsymbol k_{2}]^{T}\boldsymbol W_{1}+\boldsymbol b_{1})\\
\boldsymbol k_{mix} &= \mathrm {Relu}(\boldsymbol e_{mix}\boldsymbol W_{0}+\boldsymbol b_{0})\\
\boldsymbol k_{2} &= \mathrm {Relu}(\boldsymbol e_{2}\boldsymbol W_{0}+\boldsymbol b_{0})
\end{aligned}
\end{equation}
\vspace*{-5mm}
\subsubsection{Separated Fully-Connected Layer with Concatenation}
\vspace*{-2mm}
The last one is similar with "Share-Concat" method. $\boldsymbol e_{mix}$ and $\boldsymbol e_{2}$ are firstly input to two fully connected layers respectively, the two fully connected layers are separated, which means they do not share parameters. The output $\boldsymbol k_{mix}$ and $\boldsymbol k_{2}$ are then concatenated and input to another fully connected layer (shows in Equation \ref{separate_concat} and Figure \ref{choice} (f)). This method is further referred to ``Separate-Concat". $\boldsymbol W_{0,0} \in \mathcal {R}^{d \times d}$, $\boldsymbol b_{0,0} \in \mathcal {R}^{1 \times d}$, $\boldsymbol W_{0,1} \in \mathcal {R}^{d \times d}$, $\boldsymbol b_{0,1} \in \mathcal {R}^{1 \times d}$, $W_{1} \in \mathcal {R}^{2d \times d}$ and $\boldsymbol b_{1} \in \mathcal {R}^{1 \times d}$ are parameters of the fully connected layers.

\begin{equation}\label{separate_concat}
\begin{aligned}
\boldsymbol f(\boldsymbol e_{mix},\boldsymbol e_{2}) &= \mathrm {Relu}([\boldsymbol k_{mix},\boldsymbol k_{2}]^{T}\boldsymbol W_{2}+\boldsymbol b_{2})\\
\boldsymbol k_{mix} &= \mathrm {Relu}(\boldsymbol e_{mix}\boldsymbol W_{0,0}+\boldsymbol b_{0,0})\\
\boldsymbol k_{2} &= \mathrm {Relu}(\boldsymbol e_{2}\boldsymbol W_{0,1}+\boldsymbol b_{0,1})
\end{aligned}
\end{equation}

\begin{comment}
\begin{figure*}[h]
	\begin{minipage}{0.3\linewidth}
		\centerline{\includegraphics[width=5.5cm,height=3cm]{sub.png}}
		\centerline{(a) Sub}
	\end{minipage}
	\hfill
	\begin{minipage}{0.3\linewidth}
		\centerline{\includegraphics[width=5.5cm,height=3cm]{mul.png}}
		\centerline{(b) Mul}
	\end{minipage}
	\hfill
	\begin{minipage}{0.3\linewidth}
		\centerline{\includegraphics[width=5.5cm,height=3cm]{concat1.png}}
		\centerline{(c) Concat1}
	\end{minipage}
	\hfill
	
	\begin{minipage}{0.3\linewidth}
		\centerline{\includegraphics[width=5.5cm,height=3cm]{concat2.png}}
		\centerline{(d) Concat2}
	\end{minipage}
	\hfill
	\begin{minipage}{0.3\linewidth}
		\centerline{\includegraphics[width=5.5cm,height=3cm]{share_concat.png}}
		\centerline{(e) Share-Concat}
	\end{minipage}
	\hfill
	\begin{minipage}{0.3\linewidth}
		\centerline{\includegraphics[width=5.5cm,height=3cm]{separate_concat.png}}
		\centerline{(f) Separate-Concat}
	\end{minipage} 
	\hfill
	\caption{The training process for the four different speaker de-mixing method. (a): Sub; (b): Mul; (c): Concat1; (d): Concat2; (e) Share-Concat; (f) Separate-Concat. X-axis represents number of epochs, Y-axis represents the mean absolute error.}
	\label{train_step2}
\end{figure*}
\end{comment}

\begin{table*}[h]
	\renewcommand{\multirowsetup}{\centering}  
	\renewcommand\arraystretch{1.0}
	\setlength{\tabcolsep}{8mm}
	\centering  
	\footnotesize
	\begin{tabular}{c|c|c|c|c|c|c}
		\hline
		& \multicolumn{3}{c|}{\textbf{Cosine Similarity}} & \multicolumn{3}{c}{\textbf{Identification Accuracy (\%)}}\\
		\cline{0-6}
		\hline
		\textbf{SNR}& \textbf{-5dB} & \textbf{0dB} & \textbf{5dB}  & \textbf{-5dB} & \textbf{0dB} & \textbf{5dB}\\
		\hline
		\textbf{Before}&0.22 & 0.48&0.59 &36.5 & 58.4&72.5 \\
		\hline
		\textbf{Sub}& \textbf{0.80} & 0.82 & 0.84& \textbf{86.2}& 89.9& 95.2\\
		\textbf{Mul} & 0.68&0.73 & 0.78& 83.7& 88.8& 94.8\\
		\textbf{Concat1} & 0.44&0.47 &0.52 &52.9 &56.8 &68.8 \\
		\textbf{Concat2} & 0.51& 0.55& 0.60& 64.5& 70.3&88.5 \\
		\textbf{Share-Concat} &0.46 &0.62 & 0.69& 58.9&86.0 &92.9 \\
		\textbf{Separate-Concat}&0.78 & \textbf{0.86}&\textbf{0.89} &82.5 & \textbf{93.0}&\textbf{96.9} \\
		
		\hline
		\textbf{Clean}& \multicolumn{3}{c|}{1.0}&  \multicolumn{3}{c}{98.5}\\
		\hline
	\end{tabular}
	\vspace*{-4mm}
	\caption{The cosine similarity and speaker identification accuracy of using the estimated embedding of target speaker $\boldsymbol e_{1}^{'}$. Before denotes the cosine similarity or speaker identification directly using $\boldsymbol e_{mix}$. Clean denotes the cosine similarity or speaker identification using $\boldsymbol e_{1}$ that extracted from clean speech.}
	\label{e2-e1}
\end{table*}

\begin{table*}[h]
	\renewcommand{\multirowsetup}{\centering}  
	\renewcommand\arraystretch{1.0}
	\setlength{\tabcolsep}{8mm}
	\centering  
	\footnotesize
	\begin{tabular}{c|c|c|c|c|c|c}
		\hline
		& \multicolumn{3}{c|}{\textbf{Cosine Similarity}} & \multicolumn{3}{c}{\textbf{Identification Accuracy (\%)}}\\
		\cline{0-6}
		\hline
		\textbf{SNR}& \textbf{-5dB} & \textbf{0dB} & \textbf{5dB}  & \textbf{-5dB} & \textbf{0dB} & \textbf{5dB}\\
		\hline
		\textbf{Before} &0.60 &0.46 & 0.28&72.0 & 58.4&31.7 \\
		\hline
		\textbf{Sub} &0.78 & 0.74& \textbf{0.72}& 95.9& 90.0& \textbf{87.1}\\
		\textbf{Mul} & 0.70&0.66&0.62& 95.5& 88.4& 83.2\\
		\textbf{Concat1} &0.45 &0.42 &0.38& 65.1&56.0 & 51.7\\
		\textbf{Concat2} & 0.52&0.47 &0.42& 89.2& 70.9& 64.1\\
		\textbf{Share-Concat} &0.65&0.53&0.47 &93.7 &87.0 & 59.5\\
		\textbf{Separate-Concat}& \textbf{0.87}&\textbf{0.79}&0.70 &\textbf{97.1} & \textbf{93.8}& 83.6\\
		
		\hline
		\textbf{Clean}& \multicolumn{3}{c|}{1.0}&  \multicolumn{3}{c}{98.5}\\
		\hline
	\end{tabular}
	\vspace*{-4mm}
	\caption{The cosine similarity and speaker identification accuracy of using the estimated embedding of target speaker $\boldsymbol e_{2}^{'}$. Before denotes the cosine similarity or speaker identification directly using $\boldsymbol e_{mix}$. Clean denotes the cosine similarity or speaker identification using $\boldsymbol e_{2}$ that extracted from clean speech.}
	\label{e1-e2}
\end{table*}

\vspace*{-2mm}
\section{Experiments}\label{Experiments}
\vspace*{-3mm}
\subsection{Data}
\vspace*{-2mm}
In this work, TIMIT corpus \cite{garofolo1993darpa} is used.
The TIMIT corpus of read speech is designed to provide speech data for acoustic-phonetic studies and for the development and evaluation of automatic speech recognition systems. 
%It includes a 16-bit, 16kHz speech waveform file for each utterance.
There are a total of 6300 utterances, 10 sentences spoken by each of 630 speakers
from 8 major dialect regions of the United States. 
%70\% of the speakers are male and 30\% are female.
\begin{comment}
As two utterances of each speaker have the same word transcriptions,
they are excluded in our work to reduce possible bias. 
So there are finally 8 utterances spoken by each speaker. 
\end{comment}
The train and test set are re-split. 
Six utterances from each speaker are randomly selected for training and the other two utterances are for testing. 
Hence there are 3780 utterances in the training set and 1260 utterances in the test set.

\begin{comment}
Suppose $\boldsymbol x_{1}$ and $\boldsymbol x_{2}$ are two selected signals to combine together, $\boldsymbol x_{1}$ contains speaker 1: $\boldsymbol s_{1}$,
$\boldsymbol x_{2}$ contains speaker 2: $\boldsymbol s_{2}$, Eq \ref{add_signal} shows the combination process of the two signals, where $N$ represents the length of the signal. $\boldsymbol x_{mix}$ that contains $\boldsymbol s_{1}$ and $\boldsymbol s_{2}$ is used as the input to step two. 
\begin{equation}\label{add_signal}
\begin{aligned}
\boldsymbol x_{mix} &= \boldsymbol x_{1}+\boldsymbol x_{2}^{'}\\
\boldsymbol x_{2}^{'} &= scale * \boldsymbol x_{2}\\
scale &= \sqrt{\frac{power(\boldsymbol x_{2}^{'})}{power(\boldsymbol x_2)}}\\
power(\boldsymbol x_{2}) &= \frac{1}{N} \sum_{i=1}^{N} \boldsymbol x_{2,i}^{2}\\
power(\boldsymbol x_{2}^{'}) &= \frac{power(\boldsymbol x_{1})}{10^{SNR/10}}\\
power(\boldsymbol x_{1}) &= \frac{1}{N} \sum_{i=1}^{N} \boldsymbol x_{1,i}^{2}\\
\end{aligned}
\end{equation}
\end{comment}

In order to evaluate the performance in real world conditions, the multi-channel wall street journal audio visual corpus (MC-WSJ) \cite{lincoln2005multi} is also used in this work. MC-WSJ contains a total number of 40 speakers reading WSJ sentences in three scenarios: single speaker stationary: A single speaker reading sentences from six positions in a meeting room; Single speaker moving: a single speaker moving between six positions while reading sentences; Overlapping speakers: two speakers reading sentences from different position. There are no speaker overlap between these three conditions.

In this work, the overlapping speaker audio scenario is used. In the overlap version, there are 9 pairs of speakers containing 10 unique speakers. For each speaker pairs, there are 700 utterances in average. There are three different recording techniques: two microphone arrays, lapel and headset microphones worn on all of the speakers. 

For all of the experiments in this work, the 20 dimensional MFCC features are used \cite{snyder2018x}. 

\vspace*{-3mm}
\subsection{Experiment Setup}
\vspace*{-3mm}
For TIMIT experiments, in step one, the speaker embeddings are learned using clean TIMIT training set. After training model $\boldsymbol A$, for each speaker, 200 segments are randomly sampled and feeded into $\boldsymbol A$. The clean speaker embeddings are the average of the embeddings from each segments belonging to the same speaker. $\boldsymbol B$ is trained using the same training data as $A$. 

In step two, as TIMIT data contains clean speech only, in order to generate mixture speech signal, each utterance in TIMIT dataset are randomly mixed with another utterance from the other speaker.
More specifically, when generating mixture speech signal, one utterance contains target speaker $S_{1}$ is chosen, and an utterance from interfering speaker $\boldsymbol S_{2}$ is randomly chosen. $\boldsymbol S_{1}$ is viewed as the target speaker, and $\boldsymbol S_{2}$ is the interfering speaker. The target speaker and the interfering speaker are mixed with a certain SNR (signal-to-noise ratio). 
Training data will only be mixed with training data, test data will only be mixed with test data. This is to avoid bias problem, as when training the separation model $C$, the model will not get access to any utterances from test set.

TIMIT experiment is separated into two parts: the first one is to use $\boldsymbol e_{2}$ to obtain $\boldsymbol e_{1}$, in other words, this experiment using the embedding of the interfering speaker to obtain that of the target speaker. The second one is using $\boldsymbol e_{1}$ to obtain $\boldsymbol e_{2}$, which is using the embedding of target speaker to obtain the embedding of interfering speaker.

For MC-WSJ experiments, in step one, the speaker embeddings are learned using the headset recorded audios from the overlapping speakers scenario. The headset recorded audios are close to the corresponding speaker, as a result, the audios in this kind of recording has the close quality of the clean signal \cite{lincoln2005multi}. The same technique is used to generated and collect embedding for each speaker and training of classifier $\boldsymbol B$. 
In step two, the model $\boldsymbol C$ is trained and tested on two microphones recorded speech (microphone1 and microphone2). For each speaker pair, 70 utterances are randomly selected as the test utterances. Speaker identification accuracies are computed on this test set.

\vspace*{-5mm}
\subsection{Evaluation Metric}
\vspace*{-3mm}
In this work, two evaluation metrics are used: speaker identification accuracy and cosine similarity. 

The speaker identification accuracy is obtained from the classifier $\boldsymbol B$. After training $\boldsymbol A$, $\boldsymbol B$ is also trained and the parameters are fixed. When the de-mixing network $\boldsymbol C$ is trained, the embeddings from the test set are extracted. $\boldsymbol B$ is used to obtain the speaker identification accuracy of the test set. 

The cosine similarity score \cite{nguyen2010cosine} is directly computed between the clean embedding (e.g. $\boldsymbol e_{1}$) and the de-mixed embedding (e.g. $\boldsymbol e_{1}^{'}$). The final cosine similarity score is computed as the average of the cosine similarity scores for each sample. There is no post-processing techniques used such as PLDA \cite{kenny2013plda}, as any post-processing technique used might influence the performance the evaluation process.

\vspace*{-4mm}
\subsection{Implementation}
\vspace*{-2mm}
In this work, the dimension of all of the fully connected layers is set to 512. 
Each layer is followed by a batch normalisation layer \cite{ioffe2015batch} except for the embedding layer. ReLU activation \cite{wan2013regularization} is used for each layer except for the embedding layer. The Adam optimiser \cite{kingma2014adam} is used in training, with $\beta_1$ set to 0.95, $\beta_2$ to 0.999, and $\epsilon$ is $10^{-8}$. The initial learning rate is $10^{-3}$

\vspace*{-4mm}
\section{Results and Discussion}\label{Results and Discussion}
\vspace*{-4mm}

\begin{table}[t]
	\renewcommand{\multirowsetup}{\centering}  
	\renewcommand\arraystretch{1.0}
	\setlength{\tabcolsep}{1.1mm}
	\centering  
	\footnotesize
	\begin{tabular}{c|c|c|c|c}
		\hline
		& \multicolumn{2}{c|}{\textbf{Cosine Similarity}} & \multicolumn{2}{c}{\textbf{Identification Accuracy (\%)}}\\
		\cline{0-4}
		\hline
		&\textbf{M1}&\textbf{M2}&\textbf{M1}&\textbf{M2}\\
		\hline
		\textbf{Before}&0.46&0.41&52.1&47.1 \\
		\hline
		\textbf{Sub} & 0.74&0.69&87.2&83.9 \\
		\textbf{Mul} &0.71 &0.66&84.4& 82.1\\
		\textbf{Concat1} & 0.39&0.33 &50.2&41.7  \\
		\textbf{Concat2} & 0.64&0.60 &79.1&72.4  \\
		\textbf{Share-Concat} & 0.60&0.53 &65.1 &55.4 \\
		\textbf{Separate-Concat}& 0.83&0.80 & \textbf{91.3}&\textbf{90.9} \\
		
		\hline
		\textbf{Headset}& \multicolumn{2}{c|}{1.0}&  \multicolumn{2}{c}{99.1}\\
		\hline
	\end{tabular}
\vspace*{-3mm}
	\caption{The cosine similarity and speaker identification results on MC-WSJ dataset.}
	\label{mc_wsj_results}
\end{table}

Table \ref{e2-e1} shows the results of using $\boldsymbol e_{2}$ to obtain $\boldsymbol e_{1}$. In Table \ref{e2-e1}, the cosine similarity and speaker identification results of all of the six speaker de-mixing functions $\boldsymbol f$ in different SNR levels are shown.

Comparing without using $\boldsymbol f$ (directly evaluate on mixture embeddings $\boldsymbol e_{mix}$),
most of the architectures of $\boldsymbol f$ obtained better performance. This shows that the speaker de-mixing process removed some of the influences of the information from the interfering speakers. The ``Separate-Concat" method obtained the best performance when SNR at 0dB and 5 dB, which is close to the results of clean speech. Even the SNR is -5 dB (the power of the interfering speaker $\boldsymbol S_{2}$ is larger than the target speaker ($\boldsymbol S_{1}$), the ``Separate-Concat" method can still reach 82.5\% test accuracy and 0.78 cosine similarity. 

\begin{comment}
Figure \ref{train_step2} shows the training process of $\boldsymbol C$ of all of the six methods with SNR at 0 dB. From Figure \ref{train_step2} (f), the "Separate-Concat" method obtained lowest reconstruction loss while faster convergence. 
\end{comment}

The reason of why ``Separate-Concat'' method worked better in most of the cases might be the inputs are in different embedding spaces. $\boldsymbol e_{1}$ and $\boldsymbol e_{2}$ are pre-trained and collected, and they contain the properties of a single speaker. But $\boldsymbol e_{mix}$ contains the properties of two overlapped speakers, so it contains more complex patterns. ``Separate-Concat'' method firstly used different fully-connected layers to transform them into another embedding space, and then concatenated them. This operation might make the model to better separate different speaker properties. 

``Sub" method obtained best performance when the SNR is -5 dB. ``Sub" method, reaching 86.2\% in speaker identification and 0.80 cosine similarity score when the SNR is -5 dB. This shows that a simple mathematical operation and a linear transformation can be applied on the speaker embeddings to filter out some information of the interfering speaker. 
``Mul" method uses another mathematical operation (multiplication), and the performance obtained are still close to the that of clean speech. 

The ``Concat1", ``Concat2" and ``Share-Concat" methods obtained lower results. The reason why the ``Concat1" and ``Concat2" obtained lower performances might be because directly concatenating $\boldsymbol e_{mix}$ and $\boldsymbol e_{2}$ might influence the model $C$ to distinguish different speaker properties. The low performance of ``Share-Concat" might have the same reason.

Table \ref{e1-e2} shows the results of using $\boldsymbol e_{1}$ and $\boldsymbol e_{mix}$ to obtain $\boldsymbol e_{2}$, which is using the embedding of the target speaker to obtain the embedding of the interfering speaker. Note the SNR value is the signal-to-noise ratio of the target speaker (S1) and interference speaker (S2). So the results when SNR is 5dB is lower than the results when SNR is -5dB in Table \ref{e1-e2}. 
All of the results of six methods shows lower but close performance of that of using $\boldsymbol e_{2}$ to reconstruct $\boldsymbol e_{1}$. It shows that the ``Share-Concat" and Sub methods also have the ability to obtain high quality embedding of the interfering speaker from two-speaker environment. 

Table \ref{mc_wsj_results} shows the experiments result of microphone1 (M1) and microphone2 (M2) in MC-WSJ dataset. The ``Share-Concat" method obtain the best results, reaching 93.9\% and 90.9\% test accuracies and 0.83 and 0.80 socine similarities in M1 and M2. 
The reason why the results of M2 is lower than that of M1 might be the distance of the speakers and microphones. The M1 is closer to speakers while M2 is far from speakers \cite{lincoln2005multi}. 

Comparing with the results of headset recording, which reaches 99.1\% test accuracy, the results obtained by the ``Separate-Concat" method still have a gap. The reason might be in real world conditions, the two speakers are moving, the SNR between the target speaker and interfering speakers might be different at different time. It might be more difficult for the model to de-mix the embedding of two speakers.

\vspace*{-3mm}
\section{Conclusion and Future Work}\label{Conclusion and Future Work}
\vspace*{-3mm}
In conclusion, in this work, a speaker embedding de-mixing approach is proposed. The proposed approach reconstructs the embedding of target speaker from the embedding of interfering speaker and mixture embedding, or inversely, obtain the embedding of interfering speaker from that of target speaker and mixture embedding. The quality of embeddings are evaluated by speaker identification accuracy and cosine similarity  score on the reconstructed embeddings and the clean embeddings. Results on TIMIT (artificially augmented two-speaker signal) and MC-WSJ (real world two-speaker signal) datasets show that within the six different de-mixing architectures, the ``Share-Concat" method obtain better results, which is close to the results of clean speech. 

In this future work, more speaker mixture scenarios will be investigated, such as three-speaker mixture. 
Different model architectures might be investigated, and larger dataset might be used such as voxceleb1 and 2.

%\begin{comment}
\vspace*{-2mm}
\section{Acknowledgements}\label{Acknowledgements}
\vspace*{-2mm}
Funding for this research was provided by Huawei Innovation Research Program (HIRP).
%\end{comment}

\newpage
% -------------------------------------------------------------------------
\bibliographystyle{IEEEbib}
\bibliography{strings,refs}

\end{document}